\documentstyle[12pt,psfig]{article}
\title{\bf Star Formation Histories of the Galactic Satellites}

\author{Gerard Gilmore$^1$, Xavier Hernandez$^2$ \& David Valls-Gabaud$^3$\\
\vspace{1cm}\\
\normalsize $^1$Institute of Astronomy, Cambridge, UK\\
\normalsize $^2$Arcetri Observatory, Firenze, Italy\\
\normalsize $^3$CNRS, Observatoire Midi-Pyr\'en\'ees, Toulouse, France}
\date{\mbox{}}
\begin{document}
\maketitle
\pagestyle{empty}
%
%
\def\bull{\vrule height .9ex width .8ex depth -.1ex}
\makeatletter
\def\ps@plain{\let\@mkboth\gobbletwo
\def\@oddhead{}\def\@oddfoot{\hfil\tiny\bull\quad
``The Galactic Halo~: from Globular Clusters to Field Stars'';
35$^{\mbox{\rm th}}$ Li\`ege\ Int.\ Astroph.\ Coll., 1999\quad\bull}%
\def\@evenhead{}\let\@evenfoot\@oddfoot}
\makeatother
%
%
\def\beginrefer{\section*{References}%
\begin{quotation}\mbox{}\par}
\def\refer#1\par{{\setlength{\parindent}{-\leftmargin}\indent#1\par}}
\def\endrefer{\end{quotation}}
%
%
{\noindent\small{\bf Abstract:} 
Late accretion models for formation of the Galactic halo require that
many Galactic satellite galaxies have been cannibalised into the halo
field.  Comparison of
the metallicity and age distribution function of stars in the
surviving satellites with the apparently exclusively old stars
in the field halo can constrain the importance of any such
process. We have developed a new objective technique to determine star
formation histories in dSph galaxies. We apply this technique to the
surviving Galactic satellites, deducing an approximately uniform
distribution of ages for the constituents, quite unlike the halo field
stars. Thus, late accretion did not play a substantial part in
Galactic halo formation.
}
%
%
\section{Introduction}
 The bulk of the stellar populations in the Galactic halo field and
globular cluster stars show a well-defined turn-off, at $B-V \sim
0.4$, implying that the vast majority of the stars are old. The
fraction of stars which lie blueward of this well-defined turn-off,
with metallicities similar to that of the present dSphs, was analysed
by Unavane, Wyse, \& Gilmore (1996; hereafter UWG96) to place limits
on the importance of the recent accretion of stellar systems similar
to the extant (surviving?) dwarf satellite galaxies.  UWG96 showed
that very few ($\sim~$10 per cent) stars were found to be bluer (and
by implication, younger) than the dominant turnoff limit, with the
highest value found for the more metal-rich halo ([Fe/H]$> -1.5$).
Direct comparison of this statistic with the colour distribution
of the turnoff stars in the Carina dwarf allowed UWG96 to derive an
upper limit on the number of mergers of such satellite galaxies into
the halo of the Milky Way. This upper limit was $\sim$~60 Carina-like
galaxies. The higher metallicity data constrain satellite galaxies
like the Fornax dwarf; only $\sim~$6 of these could have been
accreted within the last $\sim~$10 Gyr. No galaxy like either
Magellanic Cloud can ever have merged. Interestingly, a limit of zero
also applies to the Sagittarius dwarf galaxy (Ibata, Gilmore \& Irwin
1995) if recent suggestions that it has a substantial membership of
near-solar metallicity are correct.

This result has recently developed even more general significance,
with comparison of the evolution in the halo formation rate in model
galaxies, calculated ab initio from hierarchical structure formation
models, with the observed evolution in the global star formation
history indicating that the latter is not inconsistent with being
largely driven by halo mergers at $z>1$. Thus one would like to be
confident that the Milky Way Galaxy really has not suffered such
mergers, before concluding that the single good test case appears not
to match the single available good model.

The question arises from the UWG96 analysis as to the validity of
adopting Carina as a `representative' dSph galaxy. Ideally, of course
one would also like to consider the age distribution function of the
dSph and the field halo stellar populations, rather than merely the
colour data. Motivated by this, and the many other scientific
applications of the method, Hernandez, Valls-Gabaud \& Gilmore (1999;
henceforth HVGG), and see also Gilmore, Hernandez, \& Valls-Gabaud
(1999), developed an objective technique to derive star formation
histories. Application of this method to the set of dSph galaxies
provides an objective determination of the age distribution in the
dSph stellar populations, for comparison with the field halo (old)
ages. This comparison has been achieved by Hernandez, Gilmore \&
Valls-Gabaud (1999, henceforth HGVG). HGVG 
used their reduction of archive HST observations of the
resolved populations of a sample of dSph galaxies (Carina, LeoI,
LeoII, Ursa Minor and Draco) uniformly taken and reduced, to recover
the star formation histories (henceforth $SFR(t)$) of each, applying
their new non-parametric maximum likelihood variational calculus
method.

\section{Objective Determination of Star Formation Histories}

A detailed description of our new method for objective determination
of star formation histories, which uses a variational calculus
approach supplemented by maximum likelihood statistical analysis, is
provided in HVGG. That reference also describes the extensive
numerical tests carried out to ensure numerical reliability in the
implementation. We provide just a short summary here.

The available and necessary information is an observed
colour-magnitude diagram, extending below the main-sequence turnoff of
the oldest population of interest, an independent determination of
the stellar metallicities, and a set of model isochrones. Given all that,
we can construct the probability that the $n$ observed stars
resulted from a certain star formation history, $SFR(t)$. This will be
given by: 
\begin{equation}
{\cal L}= \prod_{i=1}^{n} \left( 
\int_{t_0} ^{t_1} SFR(t) G_{i}(t) dt \right),
\end{equation}
where
$$
G_{i}(t)= {\rho(l_i;t) \over{\sqrt{2 \pi} \sigma(l_i)}} 
exp\left(-\left[C(l_i;t)-c_{i}\right]^2 \over {2 \sigma^2(l_i)} \right)
$$

In the above expression $\rho(l_i;t)$ is the density of points along
the isochrone of age $t$, around the luminosity of star $i$, and is
determined by an assumed IMF (the results are not sensitive to this) 
together with the duration of the
differential phase around the luminosity of star $i$.  $t_0$ and
$t_1$ are a maximum and a minimum time needed to be considered, for
example 0 and 15 Gyr. $\sigma(l_i)$ is the amplitude of the
observational errors in the colour of the stars, which are a function
of the luminosity of the stars. This function is supplied by the
particular observational sample one is analysing. Finally, $C(l_i;t)$
is the colour the observed star would actually have if it had formed
at time $t$.  HVGG refer to $G_{i}(t)$ as the likelihood matrix,
since each element represents the probability that a given star, $i$,
was actually formed at time $t$. Since the colour of a star having a
given luminosity and age can sometimes be multi-valued function, in
practice we check along a given isochrone, to find all possible
masses a given observed star might have as a function of time, and add
all contributions (mostly 1, sometimes 2 and occasionally 3) in the
same $G_{i}(t)$.  In this construction we are only considering
observational errors in the colour, and not in the luminosity of the
stars. Although the generalisation to a two dimensional error
ellipsoid is trivial, the observational errors in colour dominate
to the extent of making this refinement unnecessary.
 
The condition that ${\cal L}(SFR)$ has an extremal can be written as
$$
\delta {\cal L}(SFR)=0,
$$
and a variational calculus treatment of the problem applied.  This in
effect transforms the problem from one of searching for a function
which maximizes a product of integrals to one of solving an
integro-differential equation.  The numerical implementation required
to ensure convergence to the maximum likelihood SFR(t) is described
fully in HVGG, as are the extensive tests and simulations using
synthetic HR diagrams.

The main advantages of our method over other maximum likelihood
schemes are the totally non-parametric approach the variational
calculus treatment allows, and the efficient computational procedure.
No time consuming repeated comparisons between synthetic and
observational CMDs are necessary, as the optimal star formation
history, independent of any preconceptions or assumptions, is solved
for directly.

\section{Star Formation Histories of the dSph satellites}

Application of the variational calculus method to archival HST data
for five representative dSph galaxies has been completed by HGVG,
where further details may be found. The star formation histories of
these five galaxies cover all possible combinations. Stars in UMi are
exclusively old, with the star formation history of that galaxy
resembling that of a metal-poor Galactic globular cluster. At the
other extreme, LeoI shows continuing star formation over all times,
rising to a gentle maximum about 3 Gyr ago. Carina illustrates a more
constant rate of star formation, though again continuing over the
whole history of the Local Group. Its use as a template for the mean
star formation history is indeed justified. The CMD and derived star
formation history for Carina is shown below (Figure~1), to illustrate
the results and their diversity.

\begin{figure}
\centerline{\psfig{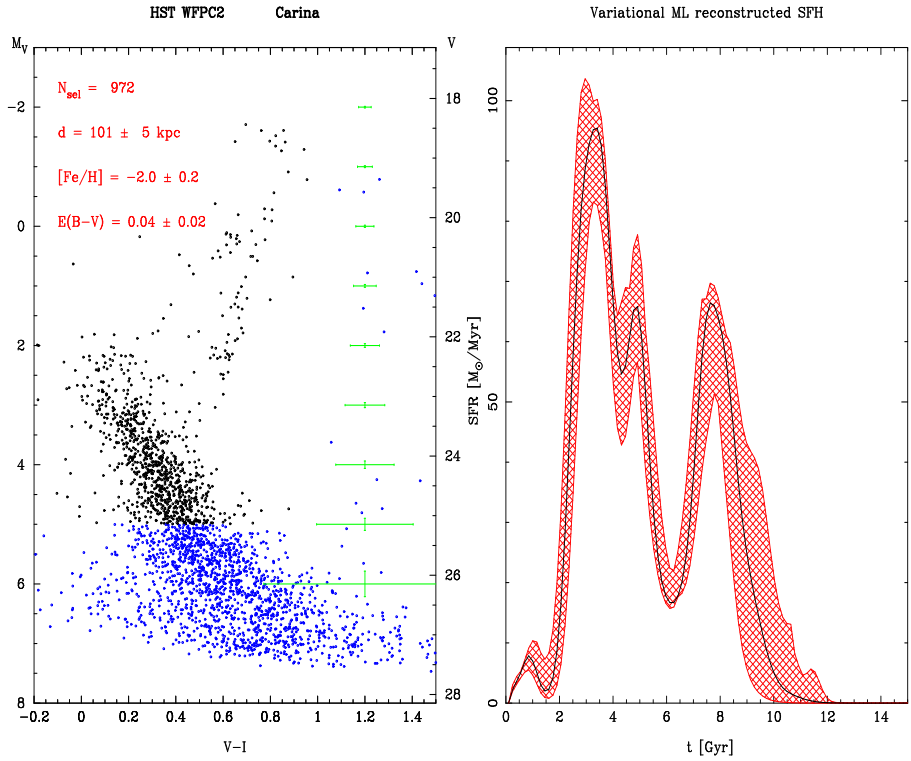}}
\caption{The HST colour-magnitude data (LHS) and corresponding star
formation history (RHS) for the Carina dSph galaxy (HGVG). This figure 
illustrates the long duration of star formation in
this dSph galaxy, an evolutionary history which is quite unlike that
of field stars in the Galactic halo.
}
\end{figure}

A special feature of the method we have developed and applied here is
that the derived star formation rate has real units, most conveniently
in solar masses per Myr, integrated over the whole dSph galaxy. Thus
it is straightorward to sum the independent star formation histories,
to provide a real history of the dSph galaxy system. This is the star
formation history of the metal-poor outer halo. The result is shown in
Figure~2. 

It is worth noting that the decline of star formation to zero
recently is an artefact of definition. Those satellite galaxies which
are still forming stars today are not dSph, but are Irregulars,
notably the LMC and SMC. Both of course are also quite metal rich, and
are very large, compared to the metal-poor galaxies of relevance here.
The rather massive Sgr dSph is also missing from this study, since
inadequate photometric and abundance data exist as yet.  

\begin{figure}
\centerline{\psfig{file=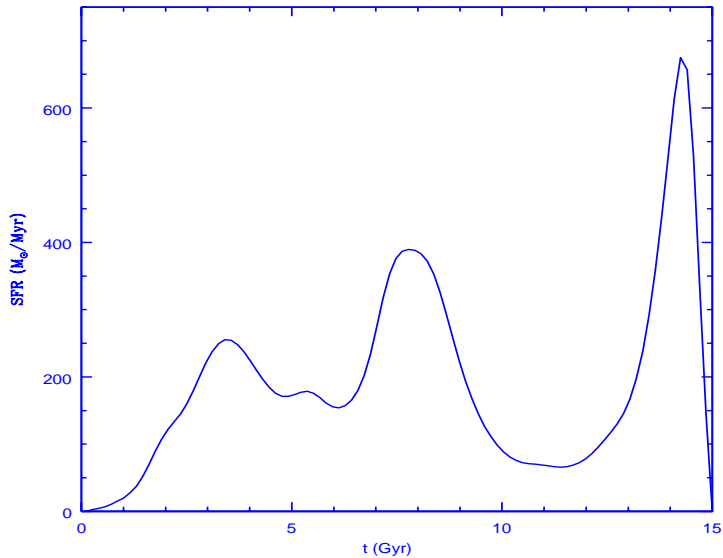,angle=0,height=8.0cm,width=10cm}}
\caption{The integrated star formation history for the five dSph
galaxies studied by HGVG. The `mean star' in a dSph satellite galaxy
is apparently of intermediate age.  }
\end{figure}

\vfill  
\section{Conclusions}
We have developed a new methodology, which removes the guesswork from
derivation of star formation histories corresponding to a given
colour-magnitude diagram. Application of this method to HST data for
five representative dSph satellite galaxies provides the star
formation history of the metal-poor Galactic satellite system. This
age distribution can then be directly compared with the equivalent
distribution for Galactic halo field stars, and globular clusters. The
Galactic field stars are, to better than 90\% accuracy, all old. Thus,
late Galactic mergers can have formed no more than some 10\% of the
field halo in the last $\sim~$10 Gyr.  This conclusion is profoundly at
variance with standard galaxy formation models, which predict early
star formation in dwarfs, which later merge to form Milky Way-sixed
spirals. Either the dwarfs merged before they formed stars, or they
never formed.

%
%
%
%
 
\beginrefer
\refer Gilmore, G., Hernandez, X,  \& Valls-Gabaud, D. 1999,
in: Astrophysical Dynamics, eds D. Berry, D. Breitschwerdt, A. da
Costa, \& J. Dyson, ApSpSci special conference issue, in press

\refer Hernandez, X,  Gilmore, G., \& Valls-Gabaud, D., 1999, MNRAS
in press (HGVG)

\refer Hernandez, X,  Valls-Gabaud, D. \& Gilmore, G., 1999, MNRAS
304, 705 (HVGG)

\refer Ibata, R., Gilmore, G., \& Irwin, M., 1995 MNRAS 277, 781

\refer Unavane, M., Wyse, R.F.G., \& Gilmore, G., 1996, MNRAS 278, 727

\endrefer           
\end{document}